\documentclass[12pt,a4paper]{article}
\usepackage{cite}
\usepackage{hyperref}
\usepackage{amsmath,amssymb}
\usepackage{braket}
\usepackage{graphicx}
\begin{document}
\title{Unitary Inequivalent Representations and Quantum Physics}
\author{Arman Stepanian\thanks{ Department of physics.University of Tehran, e-mail address: astepanian@ut.ac.ir or a89arman@gmail.com}, Mahsa Kohandel\thanks{ Department of science.Alzahra university, 
e-mail address: kohandel.mahsa@gmail.com}
}

\maketitle
\begin{abstract}
In this paper we discuss the unitary inequivalentness in quantum physics. Then based on some of the current outstanding problems in theoretical physics, we will show the important role of this concept to better understand the physical theories. 
\end{abstract}
\section{Introduction}\label{section.intro}
In Quantum Physics since we are dealing with operators on Hilbert space, it is important to construct the quantum theory in such a way that it's measurement process remains invariant under unitary transformations. From here it becomes clear that the unitarity plays an important role in quantum physics. One of the main consequences of unitary invariance (being invariant under unitary transformations) is that all physical systems which we want to study in quantum physics, should have unitary equivalent representation. The most familiar example is the wave function of non-relativistic quantum system, $\Psi(x)$. In the context of non-relativistic quantum mechanics $\Psi(x)$ is called space representation of wavefunction. Besides, we have another representation called momentum representation, denoted by $\Phi(p)$. One can transform $\Psi(x)$ and $\Phi(p)$, to each other by a fourier transformation:
\begin{equation}\label{eqn.fourier trans.1}
   \Psi(x)= \frac{1}{(2\pi\hbar)^\frac{1}{2}}\int \Phi (p)\exp \frac{ipx}{\hbar}\,dp
\end{equation}
$$
      \Phi(p)=\frac{1}{(2\pi\hbar)^\frac{1}{2}}\int \Psi(x) \exp \frac{-ipx}{\hbar}\,dx
$$
It can be easily shown that since fourier transformations preserve the norm of the wavefunction, $\Psi(x)$ and $\Phi(p)$ are unitarily equivalent.

Although in non-relativistic quantum mechanics, all representations are unitarily equivalent, different inequivalent representations are among main and natural properties of Quantum Field Theory (QFT). However, in conventional Quantum Field Theory, physicists do not pay proper attention to them. In other words in conventional Quantum Field Theory we take into consideration just one class of these representations and ignore the others. In this paper, we start with non-relativistic theory and show that in this case all physical representations are unitarily equivalent to each other. Then we discuss relativistic theories and conclude that, although these inequivalent representations play no important role in Quantum Field Theory in flat (Minkowski) space-time, they are inevitable part of Quantum Field Theory in curved space-times, without which it is impossible to formalize a consistent QFT.
\section{Equivalent Representations}\label{section.UIR}
In this section we study the equivalentness in non-relativistic quantum theory. First we make a review about the mathematical formulation of Classical Physics. Then we study the quantization procedure and finally we discuss the Stone- von Neumann theorem and it's consequences.
\subsection{Classical Physics and Symplectic Geometry}\label{Classical physics}
As we know every physical theory either classical or quantum, is formulated based on two category of objects i.e. \textbf{States} and \textbf{Observables}. In Classical Physics, the states can be regarded as the points over the phase space $\mathbb{M}$, (which is often equal to $\mathbb{R}^{2n}$). On the other hand, observables are real-valued functions from phase space to the set of real numbers $\mathbb{R}$. i.e. $f: \mathbb{M} \rightarrow \mathbb{R}$. It becomes clear then, that the measurement process in Classical Physics is just relating a real number to each point of $\mathbb{M}$ via real-valued functions.

In the Classical Mechanics, the phase space is constructed as the cotangent bundle of configuration space and it is furnished by a closed, non-degenerate and anti-symmetric map $\mathbb{\omega}$, which is known as symplectic form. So one can conclude that $\mathbb{M}$ is a symplectic manifold. The importance of this symplectic map lies on the fact that, it gives us the algebraic relations between the observables. These relations are the well-known Poisson Brackets, which is defined as follows:

\emph{Definition: For each two given functions $f(p_i, q_i, t)$ and $g(p_i, q_i, t)$ on the phase space $\mathbb{M}$ with canonical coordinates $(p_i, q_i)$
, the Poisson bracket is written as}:
\begin{equation}\label{Poisson}
\left\{f, g\right\}= \sum_{i=1}^n\frac{\partial f}{\partial q_i}\frac{\partial g}{\partial p_i}-\frac{\partial f}{\partial p_i}\frac{\partial g}{\partial q_i}
\end{equation}
where $\left\{ ,\right\}$ satisfies Distributivity, Anticommutativity and Jacobi Identity axioms.
Here it should be noticed that this definition \ref{Poisson}, comes from the non-degenerate symplectic product which is defined as:
\begin{equation}\label{Poisson 2}
 \omega (s,s')= (p_i q'_j - p'_i q_j)
\end{equation}
 where $s=(q_1,...q_n, p_1,...,p_n)$ and $s'=(q'_1,...q'_n, p'_1,...,p'_n)$ are two arbitrary points on $\mathbb{M}$ (two different states). In addition to Poisson brackets, the symplectic form helps us to find all existing coordinate transformations in $\mathbb{M}$, which leave the Poisson Bracket invariant. In terms of symplectic geometry, these kind of transformations generally are called \emph{symplectomorphisms}. However in theoretical physics they are known as \emph{canonical transformations}.
\subsection{Quantization}\label{Quant}
As we have already mentioned, \emph{states} and \emph{observables} are two main category of objects which are necessary to construct a physical theory. Once both of them are fixed, we will be able to find the mathematical and algebraic relations of them.

In the previous subsection, we have shown that Classical Physics can be considered as special case of symplectic geometry, where we have a symplectic manifold, $\mathbb{M}$, called phase space. Each point of $\mathbb{M}$ is called a state of the system, since it describes the exact position and momentum of the system. Meantime the real-valued functions from $\mathbb{M}$ to $\mathbb{R}$ are called the observables of this system. Now if we want to construct a new theory to study the physical properties of this system, we should redefine the notions of states and observables.

Quantization is a procedure which enables us to construct quantized version of a classical theory. Although there are lots of methods for quantization, Canonical Quantization is one of the most famous methods. In this method, real valued functions of the phase space are replaced by self-adjoint operators on the Hilbert space. Meantime the notion of state is achieved by replacing points and distribution functions by normalized vectors and density matrices respectively. Here normalized vectors are considered as pure states while, density matrices represent mixed states.

In addition to this, the Poisson Bracket is replaced by Canonical Commutation Relation (CCR):
\begin{equation}\label{Poisson 3}
 \left\{ ,\right\}\rightarrow i\hbar \left[ , \right]
\end{equation}

\subsection{Uniqueness Problem }\label{subsecgtion.von Neumann}

After pioneering works of founders of Quantum Theory, one question started to concern many physicists, mathematicians and philosophers:

\emph{\textbf{Is it possible to start with a classical theory, follow the quantization rules and finally arrive at two different quantized theories? }}

In other words is the quantization method gives us a unique quantized theory.
 Fortunately this question was solved by a series of papers written by M. Stone and John von Neumann \cite{Stone1930}, \cite{Neumann1931},  \cite{Neumann1932} and  \cite{Stone1932}. Their works which nowadays briefly is called  Stone- von Neumann uniqueness theorem is a mathematical theorem which states that if $\{\tilde{U}(a)| a\in\mathbb{R}\}$, $\{\tilde{V}(b)| b\in\mathbb{R}\}$ are finite sets of weakly continuous unitary operators acting irreducibly on a separable Hilbert space $H$ such that $\tilde{U}(a)\tilde{V}(b)=\exp{\frac{-iab}{\hbar}}\tilde{V}(b)\tilde{U}(a)$, $\tilde{U}(a)\tilde{U}(b)=\tilde{U}(a+b)$ and $\tilde{V}(a)\tilde{V}(b)=\tilde{V}(a+b)$, then there is a Hilbert space isomorphism $W:H\rightarrow \mathcal{L}^2 (\mathbb{R})$ such that $W\tilde{U}(a)W^{-1}=U(a)$ and $W\tilde{V}(a)W^{-1}=V(a)$.

The immediate consequence of this theorem is the fact that all quantized theories constructed via canonical quantization, from a classical theory are physically equivalent to each other. In the next subsection we discuss this consequence with more details.
\subsection{Heisenberg Group}\label{H}
 Mathematically speaking, Heisenberg group denoted as $\mathbb{H}_n$ is the $(2n-3)$ dimensional group of $n\times n$ upper triangular matrices of the form:

\[
\mathbb{H}_n
=
\begin{bmatrix}
    1 & a_1 &  \dots  & a_{n-2} & c \\
    0 & 1 &0 & \dots  & b_1 \\
    \vdots & \vdots & \vdots & \ddots & \vdots \\
    0 & 0  & \dots & 1  & b_{n-2} \\
    0 & 0 &  \dots & 0 &1
\end{bmatrix}
\]
under matrix multiplication. The elements $a_i$, $b_i$ and $c$ are the generators of $\mathbb{H}_n$ and can be taken from a commutative ring. If this ring is set to be $\mathbb{R}$ (the set of real numbers), then $\mathbb{H}_n$ is called the continuous Heisenberg group. From physical point of view this group encodes all necessary information of an n-dimensional non-relativistic quantum mechanical system, or as it has stated in \cite{Laura} \emph{\textbf{``... exponentiating the canonical commutation relations for \emph{\textbf{$\mathbb{R}^{2n}$}} yields the Heisenberg Group $\mathbb{H}_n$''}}.

The above introduced Stone- von Neumann theorem states that all irreducible representations of  $\mathbb{H}_n$ are unitarily equivalent to each other. Thus, it becomes clear that for non-relativistic quantum mechanical systems obeying Canonical Commutation Relations, the uniqueness problem is solved.

Before starting the next section we should mention that, Stone- von Neumann theorem does not guarantee the uniqueness of theories obeying Canonical Anti-commutation Relations rules (such as spin systems). However their uniqueness is proved by Wigner-Jordan theorem \cite{Jordan} (for more details see \cite{Laura}).

\section{Inequivalent Representations}\label{UIR}
As mentioned above the Von Neumann uniqueness is valid for all systems with finite degrees of freedom $N$ \cite{Haag}. But the situation changes when $N\rightarrow\infty$. In this case instead of unitary equivalent representations, we have the equivalent classes of representations. Each two representations belonging to one of these classes are unitary equivalent but the ones from different classes do not need to be equivalent. By definition the field is a system with infinite number of degrees of freedom, thus we have these inequivalent class of representations for fields. So in relativistic theories where one naturally deals with fields, the problem of uniqueness again arises. In this case since we have inequivalent representations, the fundamental question is \emph{\textbf{`` Which class of representation is the physical one?}}

 In conventional Quantum Field Theory the answer to this question is given by the condition:
 \begin{equation}\label{eqn.vacuum1}
   \hat{H}\ket0=0
 \end{equation}
 where the $\ket0$ is the vacuum state of quantum field (after this selection we ignore the existence of other representations). This selection becomes physically realizable due to existence of Poincare symmetry. As we know, conventional Quantum Field Theory is based on two physical theories namely Quantum Mechanics and Special Theory of Relativity. Thus in order to construct Poincare (Lorentz) covariant Quantum Field Theory, we have to find a unitary representation of Poincare group in Hilbert space and then conclude that the vacuum state of the field ($\ket0$) must be Poincare invariant. In other words all inertial observers (observers related to each other by a Poincare transformation) will see the same vacuum state.
Considering all above and in view that it is possible to define a globally time-like Killing vector $t$ ,we can state that:
\begin{equation}
      \exp (-i\hat{H}t)\ket0=\ket0
\end{equation}
in which $\hat{H}$ is the generator of one parameter time translation group, i.e. the Hamiltonian. 
Here we have to add that this procedure can be generalized to curved space-times when they admit a global time-like Killing vector $t$. So the vacuum state of QFT in these space-times will be invariant under the group of isometric transformations.

\subsection{Why do These Inequivalent Representations Exist?}\label{subsection.why UIR exist}
Although some people try to ignore the existence of these classes of representations, the existence of them can be proved in the context of conventional Quantum Field Theory as in Algebraic one.
In conventional approach, since we are dealing with a field, we have to make a cut for our system. Remembering that this situation does not take place in systems with finite degrees of freedom. Because in this situation we are able to close the system and specify it. But in fields this can not be done. The reason is that we can not specify the infinity. So by an idealization we make a cut and try to construct a complete set of observables locally. As an example, we suppose that our infinity is located in a very far place, say Andromeda galaxy, and ${\{A_{i}\}}$ is our complete set of observables. So it is clear that we can formalize our theory with ${\{A_{i}\}}$, but this can be done just locally. That is because if someone makes a change in a place beyond Andromeda galaxy, globally our representations will change. But as we have made a cut in Andromeda, this change will have no effect in our local observations. From one side it shows that in local observations and interactions we can neglect these different representations, but on the other side in non-local effects all of them become important. Here we have to emphasize that if we want to construct a complete and self consistent theory of Quantum Gravity which can relate local and global phenomena, dealing with all these representations is necessary.

The existence of this different representations can be easily shown in the context of Algebraic Quantum Field Theory too. Where one can associate a $ C^*-algebra $ to a quantum field. This follows by GNS construction \cite{citeulike:7477863,Segal} which states that for every element on $ C^*-algebra $ like $\omega$, there is a representation $\pi$ of the algebra by linear operators on a dense subspace $D\subseteq H$ such that
\begin{equation}\label{eqn.GNS}
\omega(A)=(\Omega,\pi(A)\Omega)
\end{equation}
where $\Omega$ is the unit vector in $D$ .
So we can conclude that in each representation which we construct with GNS construction, the specified state $\omega$ in $ C^*-algebra $ is related to unit vector $\Omega$ in Hilbert space. Thus if one chooses another state say $\nu$ and constructs another representation, then there is no need to these two representations be unitary equivalent.

Here we have to add that the existence of these representations had been realized by physicists in the early years of Quantum Field Theory due to the Schur's lemma. Moretti has stated in \cite{moretti} that every pure algebraic state $\omega$, corresponding to an irreducible representation of the algebra of observables, must inevitably select a value of $Q$ in the GNS representations.($Q$ is an observable,with arbitrary $q\in \mathbb{R}$ value on pure states) following the Schur's lemma, $\pi_{\omega}(Q)$ commutes with all elements. So it must be a multiple of identity. So two pure algebraic states $\omega$ and $\omega'$ with distinct $q$ and $q'$ ($q \neq q'$) produce inequivalent representations.

As stated above, Poincare invariance leads us to select one special class of representations. In the following subsections we intend to show that picking up one class of representations is not sufficient for describing some physical phenomena.

\subsection{Haag's Theorem}
Soon after formalization of Quantum Field Theory, physicists realized that even in the context of conventional approach to Quantum Field Theory, when we try to explain the interacting theories, more than one class of representations is needed.
The said phenomena was observed bay Haag and sometimes is called the Haag's no-go theorem, which states that \cite{HaagArt}, free and interacting fields must necessarily be defined on different unitarily inequivalent Hilbert spaces. This means that interacting Fock space cannot exist, because the frequency-splitting process cannot be applied to interacting fields. For example, it has been shown in \cite{Baker}, if we add an interacting term $\lambda \phi^4$, where $\lambda \in \mathbb{R}$ to the Klein-Gordon Lagrangian 
\begin{equation}\label{KG}
      (\Box + m^2)\phi=0
\end{equation} 
we would have 
\begin{equation}
           (\Box + m^2)\phi+\lambda \phi^4=0
\end{equation} then the mass condition $k_ak^a=m^2$
     (which plays a crucial role for frequency-splitting process) does not hold, which means that some of the single particle interacting wavefunctions will be built, in part, from plane waves with spacelike momentum vectors.
\subsection{Unruh Effect}
Beside the interacting Quantum Field Theory that indicates the natural existence of different unitary inequivalent representations, there are some important physical effects that cannot be explained based on just one class of these representations.

As mentioned above, the existence of Poincare symmetry leads us to conclude that the vacuum state $\ket{0}$ is identical for all different inertial observers.
But is it possible for non-inertial observers to see the same vacuum? 

The answer is negative. The Unruh effect is a clear example of the fact that for accelerating observer the Minkowski vacuum state $\ket0$ ( Vacuum state of a Quantum Field Theory based on Poincare symmetry) looks like a thermal state with temperature $ T= \frac{\hbar}{2\pi\kappa_B c} a$, in which $a$ stands for the acceleration. This phenomena, discovered by Unruh\cite{Unruh} , Fulling\cite{Fulling}  and Davies \cite{Davies} , plays a basic role for other important effects notably on Hawking effect. Here we discuss the Unruh effect briefly.

 As we know from Special Theory of Relativity, Rindler coordinates describes a uniformly accelerating frame of reference, which is obtained from standard Minkowski line elements: ($c=1$)
\begin{equation}
ds^2=dt^2-dx^2-dy^2-dz^2
\end{equation}
 by introducing these new coordinates:
\begin{equation}\label{transformation}
x=\xi \cosh{\eta}  
\end{equation}
$$ t=\xi \sinh{\eta}$$ 
and after some calculation we can write the so called Rindler line element: 
\begin{equation}
ds^2=\xi^2 d\eta^2-d\xi^2-dy^2-dz^2
\end{equation} where $\xi^2=\frac{1}{a^2}$. 

As stated in \cite{Arageorgis} there is a singularity at $\xi=0$. The apparent singularity at $\xi=0$ is coordinate singularity and is due to the fact that these coordinates are valid for just a portion of Minkowski space-time, called (right) Rindler wedge $R: x> |t|$.

The main feature of Rindler wedge is that it is a globally hyperbolic space-time with Cauchy surfaces $\eta=const$, where the orthogonal trajectories are $\xi^2=x^2 - t^2$. It is clear that by comparing Rindler space-time with Minkowski space-time, an observer, whose worldline is one of these hypersurfaces, undergoes constant proper acceleration of magnitude $a=\xi^{-1}$. In other words, a particle following the hyperbolic motion ,$\xi^2=x^2-t^2=const$, is a stationary observer according to Rindler coordinates.
The hyperbolicity of Rindler wedge, enables us to quantize the Klein-Gordon scalar field  $\phi$ for this space-time and to construct the Hilbert space and Fock space representations of $\phi$ with their corresponding operators. This procedure is called Fulling Quantization.

Now we may say that the Unruh effect indicates that if we consider a Klein-Gordon field equation \ref{KG}
 in the Rindler wedge and then apply the quantization procedure two times, in two different ways: one by using Minkowski coordinates and other by using Rindler coordinates we can conclude that 
\begin{equation}
\hat{N}_R\ket{0}_M\neq 0
\end{equation}
 where $\hat{N}_R$ is the number operator of Rindler quantization and $\ket{0}_M$ is the Minkowski vacuum. However, the important feature is that the two mentioned quantizations are different from each other. 

Let us look to the problem algebraically. Following Algebraic Quantum Field Theory we can say that the restriction of Minkowski vacuum state $\omega_M$ on the Rindler wedge Algebra $\mathcal{A}(R)$ defines a state $\omega_{M}\mid _{\mathcal{A}(R)}$. But contrary to Rindler vacuum state $\omega_R$, which is a pure state, $\omega_{M}\mid _{\mathcal{A}(R)}$ is a mixed one.
 It can be shown that  $\omega_{M}\mid _{\mathcal{A}(R)}$ is a KMS state \cite{Wald}:
\begin{equation}
      \rho=\prod_i \sum_{n=0}^{\infty} \exp{(\frac{-2\pi n \omega_i}{a})}\ket{n_i}_M\bra{n_i}_M
\end{equation} 
where $n\omega_i$ is the energy of $\ket{n_i}_M$ state. Thus it will become clear that $\omega_M$ can be seen as the thermal density matrix $\exp (\frac{H}{T})$, where $H$ is the Hamiltonian. Therefore  $T=\frac{a \hbar}{2\pi k_B }$.

Furthermore, the two quantizations are different in a stronger way. They are disjoint representations. That is why it is mentioned by Belinski \cite{Belinski} that \emph{\textbf{``... they refer to problems with different Hamiltonians"}}. There are several papers and useful discussions whether Unruh effect is a physical effect or it is meaningless to talk about it  \cite{Arageorgis} \cite{Belinski} \cite{Halvorson}.

\subsection{Hawking Effect}\label{section.hawking Effect}

Stephen Hawking \cite{Hawking:1974rv} showed that by regarding quantum effects, it is possible to attribute a thermal radiation to black holes. This radiation, called Hawking radiation, has a temperature which is
\begin{equation}\label{eqn.hawking.temp}
   T_{H}=\frac{\hbar c^3}{8\pi G M k_{B}}
\end{equation}
This formula contains four fundamental constants in nature, $ \hbar$, $G$, $k$ and $c $. In other words Hawking radiation showed that there is a connection between Thermodynamics, General Relativity and Quantum Field Theory. Although this was a great achievement, at first glance Hawking's original calculations suffer from transplanckian problem. Because of this, some physicists by considering the fact that in our world there is no transplanckian energy concluded that Hawking effect is not physical. Besides there was some other problems related to second law of Thermodynamics. The situation is changed when we take into consideration the Bekenstein's generalized second law of Thermodynamics \cite{Bekenstein} which states that in a system with a black hole the total amount of entropy is given by
\begin{equation}\label{eqn.bekenstein}
 \Delta S_{outside} + \Delta S_{B.H} \geq 0
\end{equation}
where for a black hole the entropy is equal to
\begin{equation}\label{eqn.entropy. beken}
 S_{B.H}=\frac{A k_B c^3}{4\hbar G}
\end{equation}
in which $A$ denotes the area of the black hole.
Since the generalized second law indicates that when an object falls into a black hole, increases the entropy of the black hole and all the information of the object will be lost, it is possible to relate a thermal radiation to the black hole. One can conclude that in the presence of gravitational collapse the vacuum state of quantum field, $ \ket {0} $, becomes unstable and finally changes to a thermal state.
The main problem arises when we try to discuss this effect in the context of Quantum field theory in curved space-time.
Consider $\ket{0}$, as a vacuum state. In order to discuss the Hawking effect we have to use the semiclassical General Relativity,
\begin{equation}\label{eqn.semiclassic}
G_{\mu\nu}=\kappa\braket{T_{\mu\nu}}
\end{equation}
where $\braket{T_{\mu\nu}}$ is the expectation value of the quantum field. Now it seems that for a vacuum state we have
\begin{equation}\label{eqn.vacuum2}
\bra{0}T_{\mu\nu}\ket{0}=0
\end{equation}.
By comparing to the left hand side of the equation, it is evident that it is compatible with the black hole metric. But the situation changes when we arrive to final state of quantum field, the thermal state $\rho$.
In this case we have
\begin{equation}\label{eqn.thermal}
\braket{T_{\mu\nu}}_{\rho}\neq 0
\end{equation}
This seems contradictory in the conventional QFT, since it is impossible to find a unitary transformation like $U$ which will be able to transform $\ket{0}$ into $\rho$ such that
\begin{equation}\label{eqn.vacuum to thermal}
U\ket{0}=\rho
\end{equation}
The strange results introduced above divide theoretical physicists into two groups. The first are those who try to resolve these contradictions using different mathematical and physical methods and concepts. Others are those who conclude that these contradictions tell us that Hawking effect is unphysical, even QFT in curved space-time does not exist.

\subsection{All Representations become equally important}
The above introduced physical phenomena, indicate that it is not sufficient to restrict ourselves to just one class of representation and try to describe all physical processes. So it seems natural that we should take into account all classes. In all cases which we have mentioned, it is possible to solve the contradictions, if we suppose that the class of representation changes to another one. Thus we should find the mathematical transformation rule which lies behind that. Although it seems to be very difficult to find the relations between inequivalent representations in QFT (like Haag's Theorem), it becomes clear and more simple when we start  to take into consideration the General Relativity. Both Unruh and Hawking effects show the complementary role of General Relativity and QFT. This means it is due to gravity and gravitational effects that one class of representation changes to another class. It is the special feature of gravitation, because neither electromagnetism nor other gauge fields of Standard Model can do that. The interesting thing is the fact that more we try to take General Relativity into account, more the mathematical relations and transformation rules  become clear. Recall the fact that in Haag's no-go theorem there is no place for General Relativity. Because we neglect the gravitational effects. Unruh effect is an intermediate case. Although it takes place in flat space-time, which means that there is no gravitational effect, there is an important difference between Unruh effect and other phenomena which we have in conventional QFT. In contrast to conventional QFT, which is formulated based on Poincare transformations and the concept of Poincare Invariance, Unruh effect is formulated based on general covariance, which is the fundamental principle of General Relativity. If we extend the notion of Poincare(Lorentz) covariant Quantum Field Theory to the general covariant one, where of course all different type of coordinate transformations are possible, it automatically leads us to the fact that in this case there is no preferred notion of states. Speaking algebraically general covariance means that for every open subset $O \in M$ there is a $\mathcal{X}  \in Diff(M)$ (Diffeomorphism group) such that  \cite{Salehi}
\begin{equation}
\pi_{\mathcal{X}}(\mathcal{A}(O))=\mathcal{A}(\mathcal{X}(O))
\end{equation}
where $\mathcal{A}(O)$ is the algebra of observables in the region $O$ and $\pi_{\mathcal{X}}$ is the representation of $\mathcal{X}$. 
Finally we arrive at the case where gravitational effects are not negligible. This is the research area which is known as QFT in curved space-times. In the next subsection we will show that if we consider all inequivalent representations, how they help us to explain some phenomena in the context of QFT in curved space-times.
\subsection{QFT in curved space-times}
One of the main examples of QFT in curved space-times, is the particle creation by gravity which has a wide range of application for many high-energy phenomena such as particle creation in expanding universe. 

\begin{figure}
  \includegraphics[width=\linewidth]{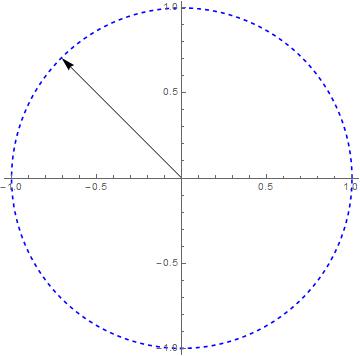}
  \caption{Each point over this unit circle can be regarded as one normalized unit vector. }
  \label{H1}
\end{figure}

\begin{figure}
  \includegraphics[width=\linewidth]{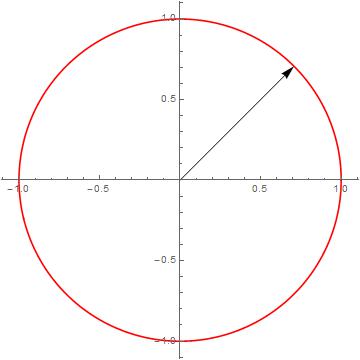}
  \caption{All unitarily equivalent representations of $\ket{0}_f$. }
  \label{H2}
\end{figure}

\begin{figure}
  \includegraphics[width=\linewidth]{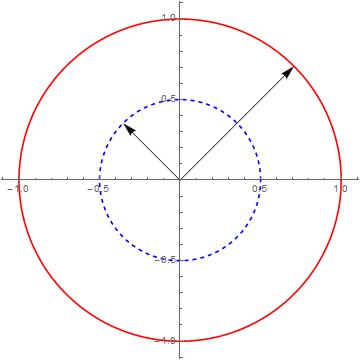}
  \caption{Equivalent representations of $\ket{0}_f$ in $\mathcal{PH}_f$ denoted as red circle and those of $\ket{0}_i$ as dashed blue. The norm of unit vectors in  $\mathcal{PH}_i$ is less than unity (here it is $\frac{1}{2}$) in  $\mathcal{PH}_f$. }
  \label{H3}
\end{figure}

Over an asymptoticaly flat space-time one can expand the field operator denoted as $\hat{\Phi}$, in terms of it's positive frequency solutions. Let $\left\{f_i\right\}$ and $\left\{F_i\right\}$ be the positive frequency solutions for past (initial time) and future (final time) respectively. Thus
\begin{equation}\label{Expansion}
\Phi=\sum_i(a_if_i+a_i^\dagger{}f_i^*)\
=\sum_i(b_iF_i+b_i^\dagger{}F_i^*)\
\end{equation}
where ($a_i$ ,$a_i^\dagger{}$) and ($b_i$ ,$b_i^\dagger{}$) are annihilation and creation operators in past and future respectively. From \ref{Expansion} it becomes clear that one can write $\left\{f_i\right\}$ in terms of $\left\{F_i\right\}$ or vice versa.
\begin{equation}\label{Expansion 2}
f_i=\sum_k(\alpha_{ik}F_i+\beta_{ik}F_i^*)\
\end{equation}
\begin{equation}\label{Expansion 2}
F_i=\sum_k(\alpha_{ik}^*f_i-\beta_{ik}f_i^*)\
\end{equation}
the $\alpha_{ik}$ and $\beta_{ik}$ are called Bogoliubov coefficients. From these definitions one can easily define the initial and final vacuum states:
\begin{equation}\label{Expansion 2}
a_i \ket{0}_i=0 \quad \textrm{and} \quad    b_i\ket{0}_f=0
\end{equation}
Meantime the number operator at final time is defined as $\mathcal{N}=b_i^\dagger{} b_i$. The fundamental statement of particle creation by gravity is the fact that, in contrast to $\ket{0}_f$ the expectation value of $\mathcal{N}=b_i^\dagger{} b_i$ for $\ket{0}_i$ is not zero.
\begin{equation}\label{Thermal}
\langle\mathcal{N}\rangle_i  = {}_i\bra{0}b_i^\dagger{} b_i\ket{0}_i = \mid \beta \mid^2
\end{equation}
where $\beta$ is defined as the sum over all ${\beta_i} $s.
Roughly speaking this effect tells us that, during the expansion of the universe pure state (vacuum state) changes to a mixed state. As we know from quantum theory, pure states can be considered as unit vectors over the Hilbert space $\mathcal{H}$ since their norm is equal to unity. The same is not true for mixed states (density matrices) because their norm is less than unity.  Now suppose that in expanding universe at some initial time $t=t_i$ we have a pure quantum state, the vacuum state $\ket {0}_i$. Figure \ref{H1}  shows this state, and it's all unitary equivalent representations on the Projective Hilbert space $\mathcal{PH}_i$. Based on conventional QFT and S-matrix theory, this vacuum state $\ket {0}_i$ can not be evolved into a mixed state. This is due to the fact that S-matrix is unitary. Now suppose that during the expansion, this class changes to another class which is not equivalent to the first class. Again at $t=t_f$ the vacuum state  $\ket {0}_f$  can be represented as a unit circle in  $\mathcal{PH}_f$ (Figure \ref{H2}). But since two representations are not equivalent, the radius of first unit circle (initial pure states) in $\mathcal{PH}_f$, is not equal to unity. Figure \ref{H3} illustrates this difference between $\mathcal{PH}_i$ and $\mathcal{PH}_f$. This change of classes (or Hilbert spaces ), becomes realizable due to gravity.   

From the  above example it becomes clear that once we move from a region with higher gravity (e.g from the early times of universe or from the vicinity of the event horizon of a black hole ) to the regions where gravity becomes weaker, the norms of Hilbert spaces where we define the pure states on them, becomes larger and larger. This is the reason that an initially defined pure state finally turns into a mixed state. This change which is caused by gravity, shows the complementary role of General Relativity and Quantum Physics. In this sense Semiclassical Gravity and even Quantum Gravity is not just en effort to quantize the Einstein's Field Equations, but also a mathematical and conceptual framework for finding all existing relations and transformation rules between the Unitary Inequivalent Representations.

\section{Conclusion}\label{section.conclusion}
As mentioned above, the existence of these inequivalent unitary representations is the inevitable part of the QFT. But the main problem is that when we try to construct a physical theory, by considering the Poincare symmetry,we select just one of these classes and simply forget about the existence of others. This causes some problems such as Haag' s no-go theorem \cite{HaagArt}. On the other hand, the formulation of S-Matrix is such that one can find the final state $|f>$ at $t=+\infty$ by operating S-matrix on the initial state $|i>$ at $t=-\infty$ without taking into account the moment of interaction, regarding it as a black box. But it is the moment of interaction that all of these classes may become equally important.

Again we have to mention that the existence of these classes is something which is related to the global structure and that is why we will not be able to see their effect  in our local observations. As stated in \cite{Wald} if $(F_1,\pi_1)$ and $ (F_2,\pi_2)$ be two unitary inequivalent representations of the Weyl Algebra $\mathcal{A}$, $A_1,A_2,...,A_n \in \mathcal{A}$ , $\epsilon_1, ... ,\epsilon_n >0$ and $\omega_1$ be an algebraic state corresponding to density matrix on $F_1$, then there exists a state $\omega_2$ corresponding to a density matrix on $F_2$ such that for all $i=1,2,...,n$ we have $|\omega_1(A_i)-\omega_2(A_i)|<\epsilon_i$ which in it's turn, shows that, although two representations of $\mathcal{A}$ may be unitarily inequivalent, the determination of a finite number of expectation values in $\mathcal{A}$, made with finite accuracy can not distinguish the difference between different representations.

Another important feature is that Quantum Gravity will enable us to relate the yet unknown transplanckian world and our one to each other. This correlation shows itself in Hawking effect which again can be explain in this manner.

Furthermore in this paper we wanted to show that there may be a new look to the yet unknown quantum theory of gravity and the gravity may have the role of relating one class of representations to another, although it seems a very difficult physical and mathematical task.
\section*{Acknowledgement}
We would like to thank Professor H.Salehi for insightful comments and valuable suggestions.

\bibliography{UnitaryInequivalentRepresentations}
\end{document}